\documentclass[%
reprint,
superscriptaddress,
nofootinbib,
amsmath,amssymb,
pra,
showkeys,
]{revtex4-2}

\usepackage{xcolor}
\usepackage[T2A]{fontenc}			 % Кодировка создаваемого файла
\usepackage[utf8]{inputenc}			 % Кодировка текстового файла
\usepackage{csquotes}
\usepackage{tempora} % Красивый шрифт типа Times

\usepackage[english]{babel} % Региональные особенности форматирования
\usepackage{ulem}					 % Зачёркивание \sout, подч. \ul и др.
\usepackage{graphicx}% Include figure files
\usepackage{dcolumn}% Align table columns on decimal point
\usepackage{bm}% bold math
\usepackage[unicode=true,colorlinks=true,citecolor=blue,urlcolor=blue]{hyperref}

\renewcommand{\i}{{\rm i}}
\renewcommand{\d}{\mathrm d}

\usepackage{braket}

\renewcommand{\d}{\mathrm d}
\renewcommand{\i}{{\rm i}}

\makeatletter
\newcommand{\make@circled}[2]{%
 \ooalign{$\m@th#1\smallbigcirc{#1}$\cr\hidewidth$\m@th#1#2$\hidewidth\cr}%
}
\newcommand{\smallbigcirc}[1]{%
 \vcenter{\hbox{\scalebox{1.3}{$\m@th#1\bigcirc$}}}%
}
\newcommand{\ohplus}{\mathbin{\mathpalette\make@circled{ \scalebox{0.7}{\textbf{D}$^+$} } } }
\newcommand{\ohminus}{\mathbin{\mathpalette\make@circled{ \scalebox{0.7}{\textbf{D}$^-$} } } }
\newcommand{\ohpm}{\mathbin{\mathpalette\make@circled{ \scalebox{0.7}{\textbf{D}$^\pm$} } } }
\newcommand{\oA}{\mathbin{\mathpalette\make@circled{ \scalebox{0.9}{\textbf{A}} } } }
\newcommand{\oB}{\mathbin{\mathpalette\make@circled{ \scalebox{0.9}{\textbf{B}} } } }
\newcommand{\oC}{\mathbin{\mathpalette\make@circled{ \scalebox{0.9}{\textbf{C}} } } }
\makeatother

\usepackage{ulem}

\definecolor{darkgreen}{RGB}{20,140,10}
\definecolor{darkred}{RGB}{190,20,15}
\definecolor{darkblue}{RGB}{0,20,150}
\definecolor{darkyellow}{RGB}{128,128,0}
\definecolor{darkviolet}{RGB}{90,0,100}
\newcommand{\addIR}[1]{\textcolor{black}{#1}}

\begin{document}

\author{V. O. Kozlov}
\affiliation{Saint Petersburg State University, Saint Petersburg 199034, Russian Federation}

\author{I. A. Smirnov}
\affiliation{Saint Petersburg State University, Saint Petersburg 199034, Russian Federation}

\author{M.~S.~Kuznetsova}
\affiliation{Saint Petersburg State University, Saint Petersburg 199034, Russian Federation}

\author{E.~V.~Kolobkova}
\affiliation{ITMO University, Saint Petersburg 197101, Russian Federation}
\affiliation{Saint Petersburg State Institute of Technology, Saint Petersburg 190013, Russian Federation}
	
\author{G.~G.~Kozlov}
\affiliation{Saint Petersburg State University, Saint Petersburg 199034, Russian Federation}

\author{V.~S.~Zapasskii}
\affiliation{Saint Petersburg State University, Saint Petersburg 199034, Russian Federation}

\author{D.~S.~Smirnov}
\affiliation{Ioffe Institute, Saint Petersburg 194021, Russia}
\affiliation{Saint Petersburg State University, Saint Petersburg 199034, Russian Federation}

\author{I.~I.~Ryzhov}
\email{i.ryzhov@spbu.ru}
\affiliation{Saint Petersburg State University, Saint Petersburg 199034, Russian Federation}

\title[Spin noise reveals spin dynamics and recharging of lead halide perovskite nanocrystals]
 {Spin Noise Reveals Spin Dynamics and Recharging of Lead Halide Perovskite Nanocrystals}

\begin{abstract}

The lead halide perovskite nanocrystals embedded into a glass matrix exhibit strong interaction with light and demonstrate exceptional optical and spin related features along with long-term chemical and physical stability. We apply the spin noise spectroscopy technique which offers a number of specific opportunities to study the spin system of CsPbI$_3$ nanocrystals in a fluorophosphate glass matrix. A pronounced spin precession peak with an isotropic $g$-factor \addIR{absolute value} of 2.7 and record dephasing time of T$_{2\text{,e}}$\,=\,2.7 ns is ascribed to resident electrons in the perovskite nanocrystals. The experimentally observed Faraday rotation noise with no noise of ellipticity is explained by saturation of the inhomogeneously broadened optical transition.
Increasing the probe intensity, we \addIR{went beyond the non-perturbative regime and observed} a number of light-induced effects. In particular, the illumination with shorter wavelength light gives rise to a persistent recharging of the quantum dots by holes ($|g|=0.17$ and T$^*_{2\text{,h}}$\,=\,1.4~ns, T$^*_{1\text{,h}}$\,$\geq$\,30~ns), which remains stable over multiple cycles of heating to the room temperature and cooling. In addition, elliptically polarized light induced an ``optical'' magnetic field on the system due to the AC Stark effect. It is confirmed using a new modification of polarization noise spectroscopy with a small degree of circular polarization of the probe light added with different frequencies.

\end{abstract}


	\maketitle

\section{Introduction}
The most valuable features of materials in the up-to-date photonics and optoelectronics are provided by their spin systems which are sensitive to the light polarization and magnetic field, can be easily excited and controlled by the light, and may exhibit extremely fast response~\cite{kramer-adv-solid-state01,bayliss-opt-addr-mol-spins20,farshchi-opt-commun-spin-info11}. Numerous effects of spin-photon interactions are widely used at present both for fundamental research and for applications in optical information processing~\cite{dey-state-of-the-art-prospects-halide-perovskite-nc21,mu-recent-progress-halide-perovskites-nc-optoel22,guan-perovskite-qds-emb-paper-photodetectors24}. Classical experiments of this kind imply measuring optical response of the medium to external perturbation~-- electric or magnetic field, microwave power, acoustic waves, etc. This class of experiments also includes the effects of nonlinear optics, when optical properties of the spin-system are perturbed by light (as, e.g., in the pump-probe spectroscopy). 

In the past decades, a conceptually new method of spin-system investigations has been developed, when properties of the spin-system are revealed not in its response to external perturbation, but rather in spontaneous fluctuations of its magnetization~\cite{zapasskii-sns-review13,kozlov-development-laser-spectroscopy-spin-noise24}. The magnetization noise of the medium, in this method, is usually detected as a noise of the Faraday rotation (FR) of the probe beam. Viability of this experimental approach, referred nowadays to as Spin Noise Spectroscopy (SNS), was primarily demonstrated on atomic systems~\cite{aleksandrov-mr-fr-noise-en81}. The SNS is currently applied most widely and efficiently to semiconductor systems~-- both bulk~\cite{romer-sns-review07,crooker-sn-gaas2009,cronenberger-atomic-like-sn15} and low-dimensional~\cite{noise-CPT,poltavtsev-sns-single-qw14,sinitsyn-spin-valley-noise-2d14,kamenskii-amplif-scatt-sn-qd2020}. 
The SNS, which may be considered as a specific method of the magnetic resonance detection, in contrast to the conventional EPR technique, does not imply real excitation of spin precession and, thus, is virtually nonperturbative. It was recently found that the SNS is also applicable to dielectric crystals with rare-earth impurities~\cite{kamenskii-re-sns2020,kozlov-sn-magn-anisotropic-centers23} and, what is more curious, to optically anisotropic crystals, when measurements of the Faraday rotation proper are hindered~\cite{kozlov-sn-birefringent22}.

A considerable attention of researchers has been attracted to the semiconductor lead halide perovskite crystals whose unique photophysical properties favorably combine with a low-cost technique of their growth~\cite{dirin-solution-grown-perovskite16,selivanov-counterdiffusion-in-gel22}. A great amount of research performed to date on these systems, in the form of both bulk and nanocrystalline samples, has confirmed their remarkable characteristics. Wide applicability of these materials in optoelectronics and solar-energy-conversion technologies, however, proved to be substantially restricted by their low chemical and mechanical stability and toxicity of Pb. To reduce the role of these factors, it was proposed to turn to a morphologically different material, namely, to nanocrystals embedded into a glass matrix~\cite{chen-robust-perovskite-qd18,chen-perovskite-qd-glasses18,kolobkova-perovskite-nanocrystals21,sixing-perovskite-qd-glasses22}. These structures, being optically homogeneous, chemically and mechanically stable and easy-to-handle, retained most merits of the bulk halide perovskites. The samples showed a bright compositionally tunable recombination emission and were amenable to optically driven spin control~\cite{belykh-coherent-spin-dynamics-e-h-perovskite19,kirstein-mode-locking-hole-spin-perovkite-nanocrystals23}. Recently, by combining the SNS abilities in application to semiconductor structures and birefringent materials, it became possible to non-perturbatively investigate the spin system of lead halide MAPbI$_3$ single crystal~\cite{kozlov-spin-noise-halide-perovskite-arxiv23}.

In this paper, we report on the first observation of spin noise in nanocrystals, or quantum dots (QDs), of the lead halide perovskites embedded into a glass matrix. 

\section{Results and discussion}

 \subsection{Spin noise of intact QDs} \label{ssec:sn-intact}
 
 This work is aimed at investigating the spin subsystem of perovskite nanocrystals in a glass matrix by means of the SNS.
 We apply this delicate technique for studying nanocrystals of the inorganic perovskite CsPbI$_3$ \addIR{which, being embedded in glass, are} much more stable in time and to environment than the \addIR{single-crystal} MAPbI$_3$ as well as many other perovskite structures and whose bandgap is governed by the size of the nanocrystals. 
 
\begin{figure}[b]
  	\centering
  	\includegraphics[width=\linewidth]{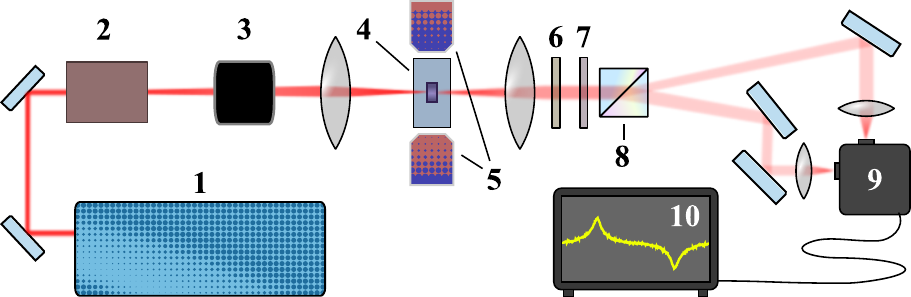}
  	\caption{Spin noise setup schematics. \textbf{1}~--- tunable cw laser, \textbf{2}~--- power stabilizer, \textbf{3}~--- Pockels cell (optional), \textbf{4}~--- sample in cryostat, \textbf{5}~--- magnet, \textbf{6}~--- half-wave plate, \textbf{7}~--- quarter-wave plate, \textbf{8}~--- polarizing beamsplitter, \textbf{9}~--- balanced photoreceiver, \textbf{10}~--- spectrum analyzer.}
  	
  	\label{fig:setup}
  \end{figure}

 The SNS experiment is basically performed as follows. A low-excessive-noise linearly polarized monochromatic light is transmitted through the sample in a cryostat at a temperature as low as 2-3~K. The wavelength of the light is tuned slightly below the absorption band to minimize optical perturbation of the spin subsystem keeping paramagnetic Faraday effect significantly high. The spins in the probed area produce a fluctuating spontaneous magnetization precessing at Larmor frequency. The magnetization precession modulates the FR of the transmitted light, and the Larmor frequency is revealed as a peak in the FR noise power spectrum. At the same time, the width of the peak equals to the inverse transverse spin relaxation time~\cite{SmirnovUFN}. The SNS experimental schematic is presented in Fig.~\ref{fig:setup}.
 
  \subsubsection{Faraday rotation noise} \label{sssec:frn}
  
  We performed the \addIR{FR noise} measurements of the glass sample with perovskite CsPbI$_3$ nanocrystals. The sample of fluorophosphate (FP) glass with the composition 35P$_2$O$_5$-35BaO-5AlF$_3$-10Ga$_2$O$_3$-10PbF$_2$-5Cs$_2$O (mol.\,\%) doped with BaI$_2$ \addIR{was} synthesized using the melt-quench technique~\cite{kolobkova-perovskite-nanocrystals21}. The glass synthesis was performed in a closed glassy carbon crucible at a temperature of T~=~1050$^\circ$C. About 50~g of the batch was melted in the crucible for 30~minutes, then the glass melt was cast on a glassy carbon plate and pressed to form a plate with a thickness of about 2~mm. Samples with a diameter of 5~cm were annealed at the temperature of 50$^\circ$C below T$_g$ = 400$^\circ$C to remove residual stresses. The CsPbI$_3$ perovskite NCs were formed from the glass melt during the quenching. The sample was polished to achieve the optical quality of the surface, and was glued with silver paste to cold finger of the closed-cycle helium optical cryostat.
   
   \paragraph{Dependence on magnetic field magnitude} \label{par:frn-field-magn}
   
   \begin{figure}
   	\centering
   	\includegraphics[width=\linewidth]{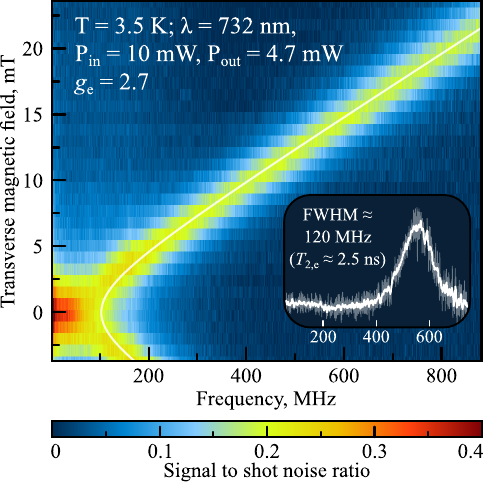}
   	\caption{Colormap \addIR{of the FR noise spectrum dependence on magnetic field. The inset shows one of the spectra obtained at minimum light power, demonstrating a relaxation rate close to the intrinsic one.}
   	}
   	\label{fig:e-sns}
   \end{figure}
   
   A~typical FR noise spectrum at 730 nm probe wavelength reveals a pronounced Larmor precession peak with a \textit{g}-factor absolute value of 2.7 (Fig.~\ref{fig:e-sns}). The peak has Lorentzian shape and shows very weak broadening with increasing field up to 25~mT (Larmor frequencies up to 1~GHz). This peak can be confidently ascribed to electrons confined in QDs, with the \textit{g}-factor \addIR{absolute} value corresponding to 1.7~eV bandgap~\cite{yang-understanding-size-dep-band-gap-perovskite-nc20,kirstein-lande-factors-e-h-lead-halide-perovskites22}. This value well correlates with the QDs average size of \addIR{15$\pm$1}~nm~\cite{meliakov-temp-dep-e-h-lande-cspbi3-nc24}. The extracted electron \textit{g}-factor spread is $\Delta g_e$\,=\,0.12 with the relative dispersion $\Delta g_e / g_e$\,=\,4.4\%. By decreasing the probe light power from 10~mW to 1.7~mW, we managed to observe $T_{2,e}\approx2.5$~ns\footnote{We speak of a homogeneous dephasing time because the peak has a well-defined Lorentzian shape.} (see the inset in~Fig.~\ref{fig:e-sns}) \addIR{Extrapolating the $T_{2,e}$ dependence on the power to the zero (not shown),} we obtain a close value of the intrinsic transverse spin relaxation time $T_{2,e}=2.7$~ns. \addIR{The dependence of the signal area on power was linear up to P$_{in}$~$\approx$~4~mW, indicating the weak perturbation of spin subsystem, and sublinear with a further increase.} At lower fields, $B \lesssim 5$ mT, the Larmor peak position deviates from linear, and the zero \addIR{frequency} peak arises thus indicating the presence of an additional effective magnetic field with its direction different from the direction of the external field. The nature of this field is discussed in the following~(\ref{sssec:nuclear-bath}).

   \paragraph{Dependence on magnetic field direction} \label{par:frn-field-dir}
   
   \addIR{Small $\Delta g_e$}
   suggests simultaneously smallness of the $g$-factor spread over the ensemble of nanocrystals (magnetic homogeneity of the ensemble) and isotropy of the $g$-factors (magnetic isotropy of individual nanocrystals).
   There is no doubt that our samples comprised of randomly aligned nanocrystals should be macroscopically isotropic. Still, we performed measurements of orientation dependence of the spin noise spectra that may provide additional and more sensitive information about hidden magnetic anisotropy of the system\addIR{, e.~g., reveal the Larmor peak presence in a Faraday geometry of the setup~\cite{kozlov-sn-magn-anisotropic-centers23}.}
   
\begin{figure}[b]
   	\centering
   	\includegraphics[width=\linewidth]{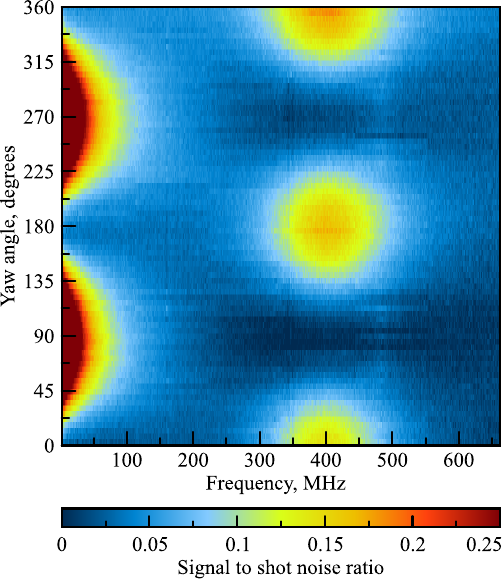}
   	\caption{The spin noise spectrum dependence on the angle of magnetic field (``yaw'', notation after Ref.~\cite{kozlov-spin-noise-halide-perovskite-arxiv23}); 0, 180, 360 degrees correspond to the Voigt geometry; 90, 270 degrees correspond to the Faraday geometry. The measurements are performed at T~=~3.5~K, $\lambda$~=~728.5~nm, and P$_\text{in}$~=~10~mW. \label{fig:angular}}
   \end{figure}

   The applied magnetic field, in these experiments, was rotated around the axis perpendicular to the light beam propagation, so that each 90$^\circ$ of rotation the Faraday and Voigt geometries interchanged. The dependence thus obtained (Fig.~\ref{fig:angular}) well corresponds to the case of isotropic medium with isotropic paramagnetic entities. In particular, there is no Larmor frequency signal in the Faraday geometry (90$^\circ$ and 270$^\circ$).
   \addIR{We thus conclude that} the birefringence of the CsPbI$_3$ crystal \addIR{itself} does not affect the polarization of the probe light nor its fluctuations, because the size of the nanocrystals is much smaller than the probe wavelength which, in turn, is significantly smaller than the length of the polarization beats. \addIR{The observed isotropy of the system, both in terms of magnetic homogeneity of the ensemble and magnetic isotropy of individual QDs,} well agrees with the results of other researchers~\cite{belykh-coherent-spin-dynamics-e-h-perovskite19,meliakov-temp-dep-e-h-lande-cspbi3-nc24}.
   
   It should be mentioned that in some cases we observed a weak flat background in the noise spectrum which was insensitive to the external magnetic field. We ascribe this component of the signal to the structural dynamics typical for glasses~\cite{kozlov-spont-noise-birefringence-re-glasses23}, which leads to stochastic linear birefringence at the excitonic resonance. \addIR{In all the data presented, this background was either subtracted or was negligibly small.}

   \paragraph{Dependence on probe power and wavelength, and sample temperature} \label{par:frn-probe}
   
\begin{figure*}
   	\centering
   	\includegraphics[width=\linewidth]{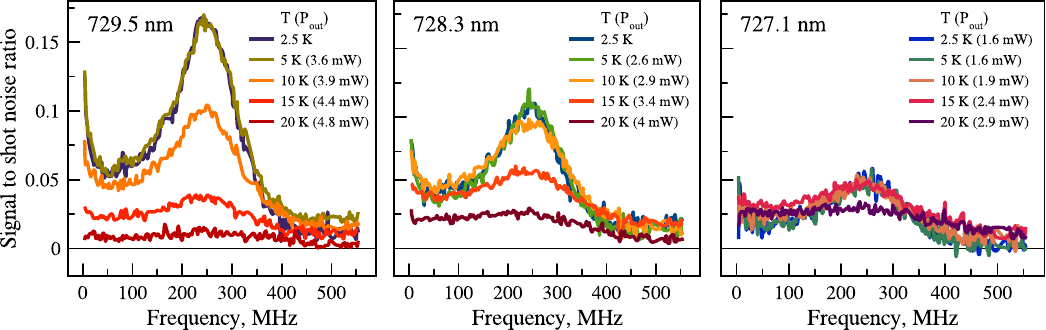}
   	\caption{Temperature dependence of the FR noise spectra for three probe wavelengths. The input power P$_\text{in}=10$~mW, the output power P$_\text{out}$ and other experimental parameters are given in the legends.}
   	\label{fig:temperature}
   \end{figure*}

   The FR noise signal remains clearly observable in the wavelength range 727--732 nm (several examples are shown in Fig.~\ref{fig:temperature}). At shorter wavelengths, the measurements are hindered by absorption, while at longer wavelengths the signal \addIR{decreases due to weakening of Faraday rotation and becomes} obscured by the light shot noise. The temperature dependence of the signal shape varies across the probe wavelength. At longer wavelengths, the Larmor peak decreases in amplitude with no width variation, what is presumably caused by the thermal shift of the bandgap and consequent increase of the detuning of the probe light. When the probe wavelength is tuned closer to the interband absorption, the peak rapidly broadens with temperature, which may be related to activation of the phonon-assisted hopping and spin relaxation in the states with weaker localization.

  \subsubsection{Ellipticity noise} \label{sssec:en}
  
  No ellipticity noise was observed in the whole range of probe wavelengths, powers, and temperatures in this geometry. This behavior stands in stark contrast to all previous spin noise measurements and, in particular, to the bulk perovskite semiconductor MAPbI$_3$, where the FR and ellipticity noise signals were identical %the FR and ellipticity noise signals were ``intermixed'' due to the birefringence of the investigated material and were indistinguishable
  under all experimental conditions~\cite{kozlov-spin-noise-halide-perovskite-arxiv23}.
  
  This fact, as we believe, is related to the strong localization of the charge carriers and to a high power of the probe light. Indeed, only for a loosely focused beam we were able to observe a very weak ellipticity noise signal.
  
  Theoretically, this situation can be described using the mechanism of the FR due to the polarization dependent optical resonance frequency shift~\cite{Nuclear_Faraday,NuclearNoise,PhysRevB.101.235416}. For a single excitonic resonance weakly coupled to the resident electron spin, we find that the FR and ellipticity signals are proportional, respectively, to the imaginary and real parts of the following expression:
  \begin{equation}
  	\mathcal E+\i\mathcal F\propto\frac{-\i V\left(2\gamma^2-2\i\gamma\delta-V^2\right)}{\left(\gamma^2+\delta^2+2V^2\right)\left(2\gamma^2+2\i\gamma\delta+V^2\right)},
  \end{equation}
  Here, $\delta=\omega-\omega_0$ is the detuning of the probe frequency $\omega$ from the optical resonance frequency $\omega_0$, $\gamma$ is the homogeneous width of the resonance, and $V$ is the optical transition matrix element (in units of frequency) proportional to the probe light amplitude. In the ensemble of nanocrystals, polydispersity of the particle sizes gives rise to significant inhomogeneous broadening, so that intensities of the FR and ellipticity noise are proportional to $\left<\mathcal F\right>^2=\int\mathcal F^2\d\omega_0$ and $\left<\mathcal E\right>^2=\int\mathcal E^2\d\omega_0$. For a high probe power, we find that $\left<\mathcal E\right>^2/\left<\mathcal F\right>^2\propto(\gamma/V)^6$ and, in particular, it reaches $1/100$ for $V/\gamma=3.7$ already. Qualitatively, relative suppression of the ellipticity noise is related to the saturation of the absorption, which is responsible for the ellipticity signal.
  
  Typically, the homogeneous linewidth is of the order of the inverse exciton lifetime $\gamma$~$\sim$~1~ns~$^{-1}$ (\cite{doi:10.1021/acs.nanolett.6b02874,PhysRevLett.119.026401,https://doi.org/10.1002/adfm.201800945}). The optical transition matrix element is determined by the power of the probe beam $P$ and matrix element of the optical transition moment $p_{\rm c.v.}$:
  \begin{equation}
  	V=\frac{2\sqrt{2}ep_{\rm{c.v.}}}{\hbar m_0\omega_0d}\sqrt{\frac{P}{c}},
  \end{equation}
  where $m_0$ is the free electron mass and $d$ is the size of the probe spot. For the experimental parameters $P=10$~mW, $d=20~\mu$m, $\hbar\omega_0=1.7$~eV, and $\hbar p_{\rm c.v.}/m_0=6.8$~eV$\cdot$\AA~\cite{kirstein-lande-factors-e-h-lead-halide-perovskites22} (this value agrees well with the exciton's radiative lifetime of about $1$~ns), we obtain $V\sim50$~ns$^{-1}$, which is indeed significantly larger than the homogeneous linewidth. Thus, we conclude that the absence of the ellipticity noise is related to the saturation of the optical resonance. At the same time, it only weakly perturbs the spin system of resident electrons, because of their weak interaction \addIR{with each other.} For example, with a 6-fold increase of the light power (from 1.7 to 10~mW), the dephasing time shortens only by a factor of 1.5 (from 2.5 to 1.6~ns).
  
 \subsection{Spin noise in a perturbative regime} \label{ssec:sn-perturb}
 
 \addIR{By varying the power, wavelength, and polarization state of the beam passing through the sample, we could make the probe light partly pumping, thereby observing interesting light-induced effects in a single-beam experiment.}
 
  \subsubsection{The effect of the probe light ellipticity} \label{sssec:ellipt}

\begin{figure}[b]
      	\centering
      	\includegraphics[width=\linewidth]{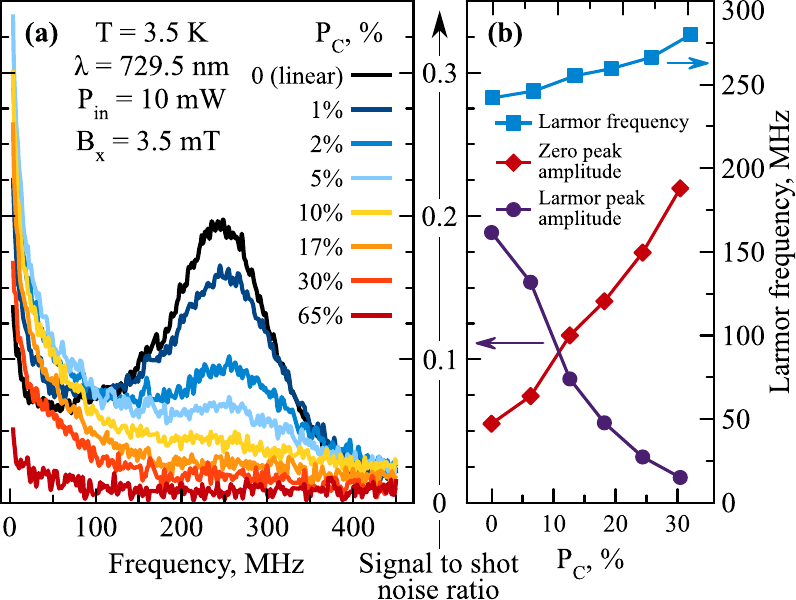}
      	\caption{\addIR{The effect of ``optical field'', caused by ellipticity of the probe light. (a) FR noise spectra modification with increase of ellipticity P$_\text{C}$, the experimental parameters are given in the legend. (b) The extracted values of the Larmor and zero frequency peak amplitudes and the Larmor peak position shift. Peak amplitudes are normalized to the total signal to account for the decrease in detection sensitivity with increasing probe circularity.} \label{fig:optical-field}}
      \end{figure}

  We performed measurements of the FR noise spectra while varying the ellipticity P$_\text{C}$ of the probe light\footnote{The degree of circular polarization is defined as \addIR{P$_\text{C}~=~2\frac{\sqrt{\text{P}_\text{min} \text{P}_\text{max}}}{\text{P}_\text{min}+\text{P}_\text{max}}$}, where P$_\text{min}$ \,(P$_\text{max}$) is the minimum (maximum) measured power after the polarizer placed into the elliptically polarized beam, notation after~\cite{ryzhov-sn-explores-local-fields16}.}, see Fig.~\ref{fig:optical-field}. We have found that with increasing ellipticity, firstly, the Larmor peak decreases and shifts, and secondly, the zero frequency peak, which was negligible before, increases strongly. This behavior indicates the appearance of an additional light-induced magnetic field, which can arise from two sources. First, it may be related to the spin polarization of nuclei and, second, it may be caused by the AC Stark effect~\cite{ryzhov-sn-explores-local-fields16}. Tentatively, we stick to the latter \addIR{due to several features of the resulting effective magnetic field.}

   \paragraph{Properties of optically induced magnetic field}

   \begin{figure}
   	\centering
   	\includegraphics[width=\linewidth]{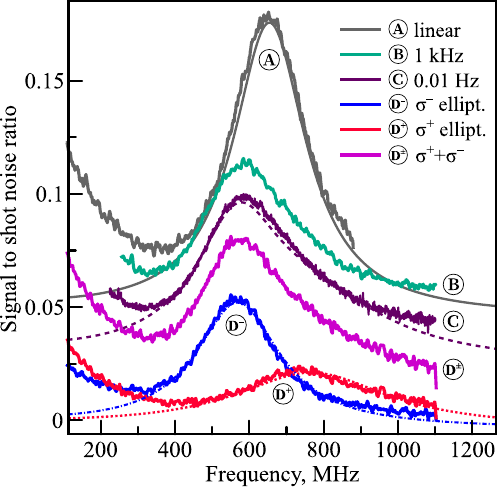}
   	\caption{\addIR{Measurements of spin noise spectra in a tilted field with elliptical probe polarization and its modulation.} B~$\approx17$~mT, P$_\text{in}= 26$\,mW, $\lambda=$\,729.6 nm, T~=~3.5~K. \addIR{The detailed description of the experimental parameters for each curve is given in the main text. The curves are shifted along the ordinate axis for ease of perception.}} \label{fig:ellipt-mod}
   \end{figure}

   \addIR{We attribute the observed phenomenon to the AC Stark effect based on the} following reasons:
   \begin{itemize}
   	\item The direction of the additional field coincided with the probe beam axis. This was confirmed by measurements in a tilted field and by geometrical calculations based on the observed peak positions and amplitudes (see Fig.~\ref{fig:ellipt-mod}, the details are \addIR{discussed} below). In the case of the dynamic nuclear spin polarization, the Overhauser field is parallel to the external magnetic field~\cite{OptOr}.
   	\item Despite the relatively low power density in these experiments ($\sim$1~kW/cm$^2$ compared to~$\sim$300~kW/cm$^2$ in \textit{n}-GaAs microcavity~\cite{ryzhov-sn-explores-local-fields16}), it is definitely beyond the non-perturbative regime, what is confirmed by the broadening of the spin noise spectrum and absence of the ellipticity noise, as discussed above.
   	\item The characteristic time of the additional field uprise was shoter than 1~s (minimal acquisition time for a single spectrum). The nuclear polarization relaxation times are expected to be longer~\cite{kirstein-lead-dom-hyperfine-int-impact-spin-dyn-halide-perovskites21}.
   \end{itemize}
   
   \paragraph{SNS modification with ellipticity modulation}
   
   Still, since 1~s time resolution can be too rough to resolve fast nuclear polarization effects~\cite{kirstein2023squeezed}, we developed a modification of the \addIR{SNS} measurement protocol to achieve 1~ms resolution. To do so, we placed a Pockels cell into a probe beam (shown in Fig.~\ref{fig:setup} as an option), which allowed us to modulate the light ellipticity at frequencies up to several kHz. After that, we tuned the parameters of the measurement to achieve the clearly observable Larmor peak shifts of opposite signs for clockwise and counter-clockwise elliptically polarized beams. This was achieved by applying a tilted field of $\sim$17~mT, which contained components both parallel and perpendicular to the light beam, and probed the sample with the elliptical light switching between $\sigma^+$ and $\sigma^-$.
   
   In Fig.~\ref{fig:ellipt-mod} the $\oA$ spectrum represents the initial unperturbed signal with Lorentzian shape (solid line fit) acquired with linearly polarized light (P$_\text{C} < 1\%$). The curves $\oB$ and $\oC$ represent the measured spectra with a meander $\sigma^+$/$\sigma^-$ modulation (P$_\text{C} \approx 48\%$) at frequencies 0.01~Hz and 1000~Hz, respectively. The frequency was scanned with 5 times multiplication per step, intermediate spectra are omitted.
   
   In addition, we performed the measurements with fixed $\sigma^+$ and $\sigma^-$ elliptical polarizations, plotted in Fig.~\ref{fig:ellipt-mod} as $\ohplus$ and $\ohminus$, respectively (they are halved for illustrative purposes). The curve $\ohpm$ shows the sum of~$\ohplus$~and~$\ohminus$. The Larmor peaks in $\ohplus$~and~$\ohminus$ are fitted with Lorentzians (dotted and dash-dotted curves, respectively). The $\oC$ spectrum is compared with a sum of unmodified $\ohplus$~and~$\ohminus$ approximations, which demonstrates a very good agreement. It is seen from Fig.~\ref{fig:ellipt-mod} that in the covered frequency range spectra reveal almost no variation of the shape. We have also checked for the absence of slow dynamics on a scale of tens of minutes.

      \begin{figure}[b]
      	\centering
      	\includegraphics[width=\linewidth]{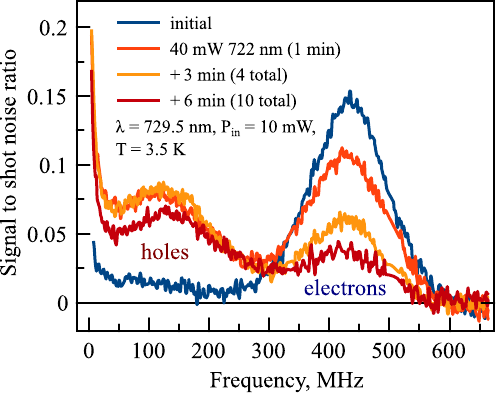}
      	\caption{Modification of the spin-noise spectrum after 720 nm light irradiation: appearance of the \addIR{holes} Larmor precession peak with $|g|\approx0.17$ and of a narrow zero-frequency peak. Simultaneously, amplitude of the electron precession peak decreases. \label{fig:irradiation}}
      \end{figure} 

   This allows us to conclude that the action of elliptically polarized light is instantaneous to within a millisecond, which ultimately makes the influence of dynamic polarization of nuclei hardly probable. Essentially, the small circular polarization of the probe beam produces spin polarization of the optical resonance equal to
   \begin{equation}
   	L_z=\frac{V^2P_c/2}{2V^2+\delta^2}
   \end{equation}
   (for $V\gg\gamma$), which creates an additional exchange magnetic field along the direction of the probe beam.

  \subsubsection{Recharging of the nanocrystals}
  
\begin{figure}
   	\centering
   	\includegraphics[width=\linewidth]{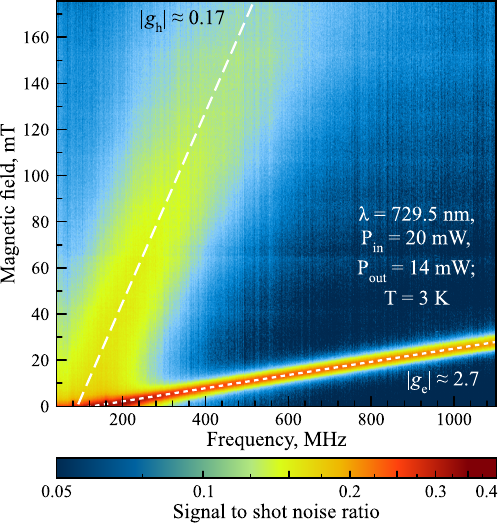}
   	\caption{Spin noise signal dependence on magnetic field after short wavelength irradiation (722 nm, 40 mW, 10 minutes). Parameters of the experiment are given in the legend. \label{fig:h-sns}}
   \end{figure}

   \paragraph{Conditions for the optical recharging}
   When studying spectral characteristics of the spin noise, we observed a curios effect of nanocrystals permanently losing the negative charge and acquiring the positive one. It was found that, in the spot preliminary pumped for several minutes by a shorter wavelength (720~nm\addIR{/}1.72~eV), a persistent spin noise peak arose with a \textit{g}-factor absolute value of around 0.17, as shown in Fig.~\ref{fig:irradiation}. This peak was not extinguishable by additional optical pumping at other wavelengths. This peak with a strongly different field dependence (Fig.~\ref{fig:h-sns}) was attributed to \addIR{resident} holes due to its $g$-factor \addIR{absolute value}~\cite{meliakov-temp-dep-e-h-lande-cspbi3-nc24}. Interestingly, it was not possible to induce the same recharging effect by shorter wavelength pumping, for example, by 650 or 532 nm laser beams. We ascribe this behavior to the fact that SNS probes the material \addIR{in its volume}, whereas the shorter-wavelength light is completely absorbed in a thin near-surface layer of the crystal, being unable to affect \addIR{it in depth}.
   %the signal that comes from bulk of the sample is not affected by this irradiation. 
   By contrast, optical density of the sample at 720 nm is essentially lower, so that the light is capable of reaching the bulk and causing the optically induced recharging in the probed volume. This assumption is additionally confirmed by weakness of the hole generation by the 730 nm irradiation \addIR{which became barely noticeable after several hours of probing the sample}.

   It is noteworthy that Fig.~\ref{fig:h-sns} was obtained after recharging, heating up to room temperature and then cooling down to 3~K on the next day. Further experiments confirmed the ultimate persistence of the induced recharging, which was not suppressed by numerous cooling cycles and did not diffuse away from the irradiated spot.

         \begin{figure}
         	\centering
         	\includegraphics[width=0.9\linewidth]{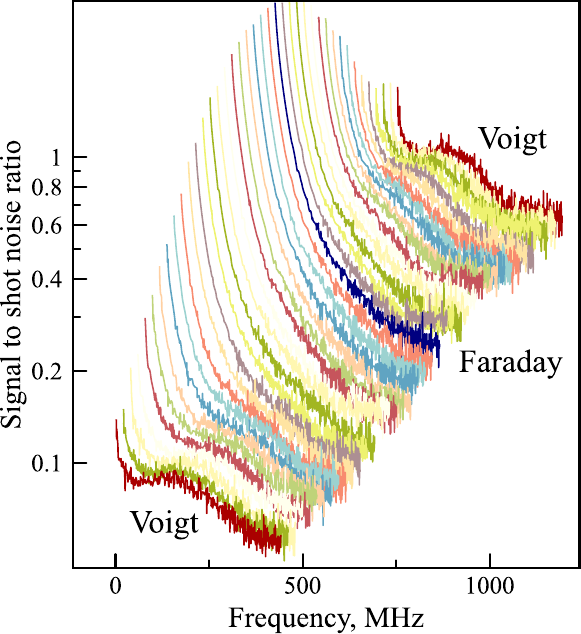}
         	\caption{The dependence of the spin noise spectrum in the \addIR{preliminary} irradiated spot on the angle of magnetic field (``yaw'', pure Voigt and Faraday geometries are accentuated). T~=~3.5~K, $\lambda$~=~728.5~nm, P$_\text{in}$~=~12.5~mW, B~$\approx$~65~mT. \label{fig:angular-h}}
         \end{figure}

   \paragraph{Characteristics of the hole ensemble}
   To further verify correctness of our observations and our conclusions, we performed the experiments described in the previous sections in the area with photoinduced holes. We obtained qualitatively similar results for the ellipticity noise measurements (no measurable signal), dependence on the magnetic field orientation (see Fig.~\ref{fig:angular-h} for details), and measurements with elliptically polarized probe (we observed a relatively small, but reproducible shift of the hole Larmor peak).
   
   At the same time, there were some differences compared to the signal without recharging. Particularly, the Larmor peak of holes experienced stronger broadening with increasing magnetic field, and its shape evolved from Lorentzian in low fields to more Gaussian-like.
   This indicates stronger inhomogeneity of $g$-factors~\footnote{For the simple structure of the top of the valence band the anisotropy of the $g$-factors is likely negligible.}: $\Delta g_h$\,=\,0.057 and $\Delta g_h / g_h$\,=\,1/3. This can be related to the formation of defects and corresponding modification of the localizing potential. \addIR{The maximum observed dephasing time was $T^*_{2,\text{h}}\,=\,1.4$~ns.}
   
   A notable feature of recharging is the appearance of the \addIR{pronounced} narrow zero-frequency peak (see \addIR{Figs.~\ref{fig:irradiation} and \ref{fig:angular-h}}\footnote{The area under spectra is not conserved, because the Larmor peak of electrons is outside the detection bandwidth, but the conservation of area was verified at lower fields.}). The amplitude of this peak increases with short wavelength irradiation proportionally to the amplitude of the Larmor peak of holes, so we ascribe this peak to longitudinal relaxation of holes. This peak is most pronounced in the Faraday geometry (Fig.~\ref{fig:angular-h}). The width of this peak, depending on the experimental conditions, can be less than 10 MHz, which evidences that the longitudinal spin relaxation time $T^*_{1\text{,h}}$ of holes can exceed 30~ns.
       
     \begin{figure}
     	\centering
     	\includegraphics[width=\linewidth]{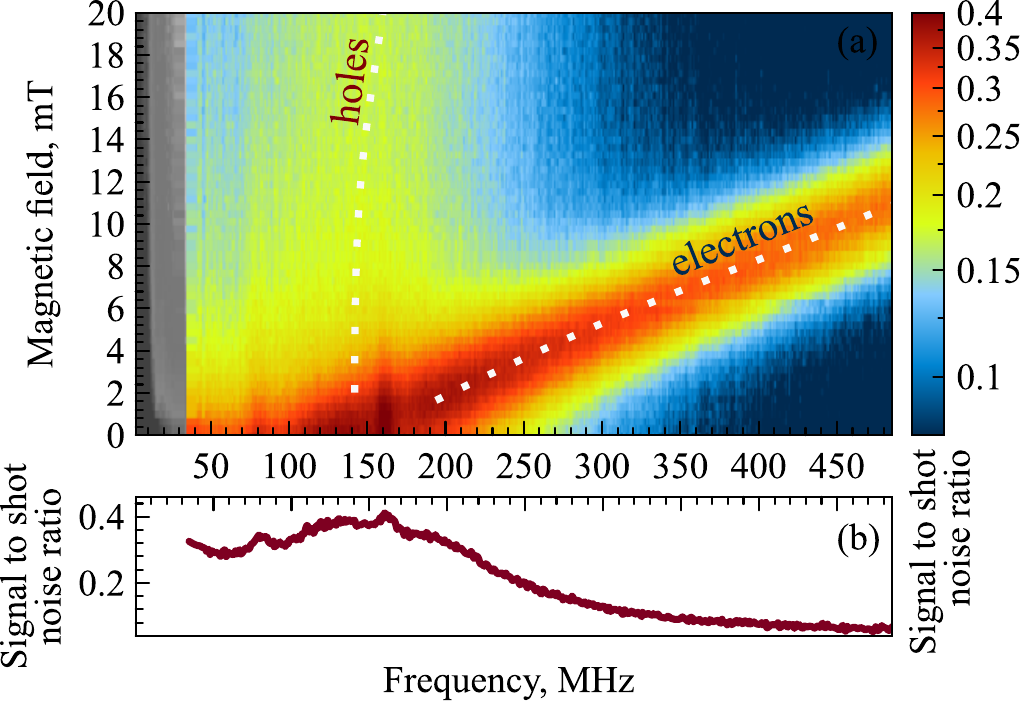}
     	\caption{(a) The detailed spin noise signal \addIR{colormap} of electrons and holes at low magnetic fields. \addIR{The grayed out area contains unsuppressed low-frequency excess noises and does not correspond to spin fluctuations. The dotted lines are the guides for the eye.} (b) The FR noise spectrum at zero magnetic field \addIR{(the bottommost cut of the colormap).} \label{fig:nuclei}}
     \end{figure} 
  
  \subsubsection{Effect of nuclear bath} \label{sssec:nuclear-bath}
  Interestingly, \addIR{the hole-related zero-frequency} narrow peak does not completely vanish in the Voigt geometry. \addIR{Similar behavior, as noted earlier~(\ref{par:frn-field-magn}), is also observed for negatively charged QDs at low magnetic fields.} This is known to be a fingerprint of the hyperfine interaction with the nuclear spin fluctuations~\cite{NoiseGlazov,PolarizedNuclei}, which is more efficient for holes than for electrons in perovskite structures~\cite{belykh-coherent-spin-dynamics-e-h-perovskite19,kirstein-lead-dom-hyperfine-int-impact-spin-dyn-halide-perovskites21}. Additionally, we found the stronger effect of the elliptically polarized probe, which may point out to the nuclear spin polarization effect in nanocrystals with photoinduced holes.
  
  To verify the effect of the nuclear bath, we have studied the region of small magnetic fields in more detail, see Fig.~\ref{fig:nuclei}. With decrease of the magnetic field down to zero, the spin precession peaks of electron and holes do not come to zero frequency, but saturate at about 150~MHz, as shown in Fig.~\ref{fig:nuclei}(b). This behavious can not be related to the residual field of magnetic cores of the electromagnet, because the electrons and holes \textit{g}-factor \addIR{absolute values} differ by 16 times. Thus we attribute this frequency to the hyperfine interaction with the nuclear spin fluctuations. Since the corresponding constant is somewhat larger for holes~\cite{meliakov-h-spin-precess-deph-ind-nuclear-hyperfine-perovskite-nc24}, the equal frequencies point out to the deeper localization of electrons as compared with holes.

\section{Conclusion}

We have investigated the lead halide perovskite CsPbI$_3$ nanocrystals (\addIR{15} nm size) embedded in a glass matrix by means of \addIR{spin noise spectroscopy}. We measured the absolute value of the $g$-factor of the resident \addIR{electrons} to be 2.7 with the spread of 4.4\%. The spin ensemble revealed magnetic isotropy in its Lorentzian shape and in the rotating magnetic field. The dephasing time \addIR{$T_{2,e}$} of electrons was found to be as long as 2.7~ns. In addition to the almost \addIR{non-perturbative} regime of SNS, we observed \addIR{several} light-induced effects: (i) Absence of the ellipticity noise, explained theoretically by the optical resonance shift mechanism of the Faraday rotation under saturation of optical transitions. (ii) AC Stark effect for elliptically polarized light, which was established using original \addIR{modification of} SNS technique with oscillating circular polarization degree providing millisecond time resolution. (ii) Persistent recharging of nanocrystals with holes under irradiation with 720~nm wavelength beam, which was to stable to heating and cooling of the sample and did not diffuse. The photoinduced holes revealed transverse and longitudinal spin relaxation times $T^*_{2\text{,h}}$\,=\,1.4~ns and $T^*_{1\text{,h}}$\,$\geq$\,30~ns, respectively, together with the $g$-factor of |\textit{g}$_\text{h}$|\,=\,0.17 and its spread of 33\%.

Thus, using the SNS we gained a deep insight into the spin physics of perovskite nanocrystals in a non-peturbative manner, locally in the \addIR{volume} of the sample. In particular, the observed effect of persistent recharging opens the way towards optically operated non-volatile memory based on perovskite nanocrystals.

\begin{acknowledgments}
	
The spin noise experiments were financially supported by Ministry of Science and Higher Education of the Russian Federation Megagrant №~075-15-2022-1112. The theoretical modelling by D.S.S. was supported by the RSF grant №~23-12-00142. The sample preparation was supported by SPbSU grant №~1024022800259\nobreakdash-7. The experimental work was fulfilled using the equipment of Resource Center ``Nanophotonics'' of Saint Petersburg State University Research Park.

\end{acknowledgments}

\bibliography{nanocrystals-SNS}

\end{document}